\documentclass[twocolumn, a4paper]{IEEEtran}

\usepackage[english]{babel}

\usepackage{amssymb}
\usepackage{amsthm}
\usepackage{amsmath}
\usepackage{graphicx}
\usepackage{dsfont}
\usepackage{float}
\usepackage{tikz}
\usepackage{mathtools}
\usepackage{physics}
\usepackage{hyperref}
\usepackage{multirow}
\usepackage{csquotes}
\usepackage{diagbox}

\usepackage[switch]{lineno}

\usepackage{tikz}
\usetikzlibrary{calc,3d,decorations.pathreplacing}

\usepackage[%
backend=bibtex8,
style=numeric,
giveninits=true,
sorting=none,
maxbibnames=99,
doi=false,
isbn=false,
url=false,
eprint=false
]{biblatex}
\newcommand{\Nunet}{U-net}
\newcommand{\Nresnet}{ResNet}
\newcommand{\Nsegnet}{SegNet}
\newcommand{\Npspnet}{PSP-net}
\newcommand{\Nmobilenet}{Mobile-net}
\newcommand{\Nvggnet}{VGG-net}
\newcommand{\Nfcnnet}{FCN-net}

\title{Encoder-decoder semantic segmentation models for
  electroluminescence images of thin-film photovoltaic modules}
\author{Evgenii Sovetkin, Elbert Jan Achterberg, Thomas Weber, and Bart E. Pieters
\thanks{Manuscript received xxx,2020}
\thanks{Evgenii Sovetkin and Bart E. Pieters are with the
        IEK5-Photovoltaik, Forschungszentrum J\"ulich, 52425 J\"ulich,
        Germany (e-mail:
        \href{mailto:e.sovetkin@fz-juelich.de}{e.sovetkin@fz-juelich.de})
        }
\thanks{Elbert Jan Achterberg is with Solar Tester, Schinnen, Netherlands}
\thanks{Thomas Weber is with PI-Berlin, Berlin, Germany}
\thanks{Digital Object Identifier (DOI):}}

\begin{document}
\linenumbers
\let\makeLineNumber\relax

\maketitle

\begin{abstract}
  We consider a series of image segmentation methods based on the deep
  neural networks in order to perform semantic segmentation of
  electroluminescence (EL) images of thin-film modules. We utilize the
  encoder-decoder deep neural network architecture. The framework
  is general such that it can easily be extended to other types of
  images (e.g.\ thermography) or solar cell technologies (e.g.\ %
  crystalline silicon modules). The networks are trained and tested on
  a sample of images from a database with 6000 EL images of Copper
  Indium Gallium Diselenide (CIGS) thin film modules. We selected two
  types of features to extract, shunts and so called ``droplets''. The
  latter feature is often observed in the set of images. Several models
  are tested using various combinations of encoder-decoder layers, and
  a procedure is proposed to select the best model.  We show exemplary
  results with the best selected model. Furthermore, we applied the
  best model to the full set of 6000 images and demonstrate that the
  automated segmentation of EL images can reveal many subtle features
  which cannot be inferred from studying a small sample of images. We
  believe these features can contribute to process optimization and
  quality control.
\end{abstract}

{\bf Keywords:} encoder-decoder neural networks, thin-film,
electroluminescence imaging

\section{Introduction}

Recently there has been an increasing interest in automated image
analysis of spatially resolved characterization methods for Photo
Voltaic (PV) modules such as electroluminescence\linelabel{typo: luminescence} (EL)
\cite{demant2018deep,
  deitsch2018segmentation, deitsch2019automatic, de2019automatic,
  karimi2019automated, sovetkin2019automatic}. Such automated image
analysis aims at quality control of modules and is thus of great
interest for manufacturers, PV system owners, and insurance companies,
as it allows for a systematic inspection of a large number of modules,
both prior and after installation.

Several commonly used PV imaging methods exists which reveal detailed
information on the state of a PV module. Examples include ultraviolet
fluorescence luminescence (UVFL) to inspect the encapsulation of a
module, Electro- and Photo-luminescence, which provides detailed
information on the local electrical properties of the solar cells, and
thermography, with which the operation temperature distribution within
a module may be estimated.

These imaging methods may all be applied prior to installing the modules
and after installation to inspect a system during operation. Thus,
manufacturers may use these methods in production quality control.
System owners and insurance companies may also use PV imaging methods
to determine the cause of module performance issues or to support
warranty claims and determine liabilities.

An automated image analysis allows the systematic analysis of a large
number of module images. Thus, it can greatly contribute to classify
degradation modes of modules and to the identification of early
warning signs or degradation.

In this work we demonstrate the application of novel semantic image
segmentation methods based on deep neural networks. The aim of an
image segmentation is to assign a label to every pixel in an image. In
our case we would like to identify pixels in EL images that correspond
to some defect in a module.

We apply our segmentation models to a database of 6000 EL images of
thin-film modules. Several common defect types are identified and
different models are trained to detect these defect. Furthermore, we
propose an approach to select the best models from the set of the
obtained trained segmentation networks.

The main challenge that we have to overcome in this work comes from
the type of our PV modules. Whereas for crystalline silicon there
exists a well established catalogue of defects visible in EL images
\cite{iea2014review}, such a catalogue does not exist for thin-film
modules\linelabel{thin-film defect catalogue}. Therefore, our methods must be flexible enough to be easily
adapted for different defects. This motivates the choice of the deep
neural network methodology that we use.

We demonstrate that our approach reliably detects various defect
types. Furthermore, a combined statistical evaluation of the EL image
database reveals hidden features in EL images that are not observed in
individual images of the modules. Our methods can be easily adapted
for other types of defects, as well as other types of technology.

The paper is organized as follows. Section~\ref{sec:related-works}
reviews literature on the subject of automatic image analysis. The
available data used in this study and its preprocessing is discussed
in Section~\ref{sec:data}. Section~\ref{sec:methods} elaborates on our
data and methodology. Section~\ref{sec:results} discusses the
results. Lastly, this work is concluded in
Section~\ref{sec:conclusions}.



\section{Related works}
\label{sec:related-works}

To review the relevant literature in a structured way we split this
section into two subsections. Firstly, we review the research on the
automatic visual inspection in photovoltaics. Secondly, we discuss
works in other application fields, where approaches similar to this
paper have been used. This allows us to overview methods used so far
in photovoltaics community as well as to capture research trends in
other image analysis fields.

One of such trends is the gradual replacement of feature-based methods
by machine learning techniques. By a feature-based method, we
understand here an approach that applies some transformation to an
image that is tailored for extraction of some particular
information. On the other hand, the machine learning methods usually
solve some generic tasks such as regression or classification, which
are then adapted to a particular application. This work also adapts a
set of generic image segmentation methods based on deep neural
networks.

\subsection{Image analysis in PV}

In photovoltaics automated image analysis methods aim to solve
different tasks. Although very different aims are pursued, there are
some common methods applied. For this reason we briefly review related
works on image analysis which can be roughly categorized as follows;
detecting and locating defects and other structures of interest,
forecasting module performance, and image collection itself. 

Most work on locating and identifying structures of interest revolves
around cracks in crystalline silicon solar modules.
\citeauthor{tsai2010micro} \cite{tsai2010micro},
\citeauthor{anwar2012micro} \cite{anwar2012micro, anwar2014micro} used
a pipeline consisting of an anisotropic diffusion and certain shape
analysis algorithms to localise cracks in EL images.

\citeauthor{tsai2012defect} \cite{tsai2012defect} consider image
representation in the Fourier domain to identify position of cracks,
breaks and finger interruptions. There the Fourier transformed image
is filtered by setting high-frequency coefficients associated with
lines artifacts to zero. Then the defects are identified by comparing
the original image and the high-pass filtered image. Due to the
assumptions on the shape of the defect, the method has difficulties
detecting defects of more complex shapes.

\citeauthor{ica_tsai2012defect} \cite{ica_tsai2012defect} introduce a
supervised learning method for defect identification that uses
Independent Component Analysis (ICA). Manually selected defect-free
solar cell subimages are used to find a set of independent basis
images with an independent component analysis (ICA), that are
consequently used in a cosine distance to determine presence of a
defect in a test sample image.

Some works focus on an automatic visual inspection of very specific
parts of a PV module. \citeauthor{SunEtAl2010} \cite{SunEtAl2010}
proposed a machine vision algorithm to examine electrical
contacts. Whereas \citeauthor{tseng2015automatic}
\cite{tseng2015automatic} describe a method for an automatic detection
of finger interruptions via binary clustering.

Infrared imaging is mostly used to detect hot-spots. As IR imaging
provides fairly direct information on the local surface temperature of
a module, relatively simple image processing algorithms can be used.
\citeauthor{chaudhary2017efficient} \cite{chaudhary2017efficient} uses
watershed transform algorithm to identify the hot-spots.
\citeauthor{ngo2016image} \cite{ngo2016image}, and
\citeauthor{alsafasfeh2017fault} \cite{alsafasfeh2017fault} use
clustering algorithms to segment IR
image and identify hotspots. \citeauthor{hepp2016irhotspot} developed a
thresholding method for hot spot detection \cite{hepp2016irhotspot}. 

There are several research directions that have been established in
order to build models that forecast electrical characteristics from
images. \citeauthor{PotthoffEtAl2010} \cite{PotthoffEtAl2010} uses a
physical model to calculate the operating voltage of individual
crystalline solar cells by EL imaging.

More recent works do not rely on a particular physical model and train a
machine learning method to extract the required information from the
data. \citeauthor{mehta2018deepsolareye} \cite{mehta2018deepsolareye}
proposed a system for forecasting power loss, localisation and type of
soiling from RGB images of solar modules. Their approach uses deep
neural network architectures similar to ones we use in this paper.

\citeauthor{deitsch2019automatic} \cite{deitsch2019automatic} train an
SVM classifier based on the extracted feature descriptors (SURF, KAZE,
FAST), and a \Nvggnet{} based neural network in order to identify
defective cells that have an impact on power reduction of the whole
module.

\citeauthor{demant2018deep} \cite{demant2018deep} proposes a
Convolution Neural Network (CNN) architecture to forecast IV
characteristics from a PL image as a production process control
procedure.

When it comes to an application of these automatic methods to the real
data, several practical problems arise. It is often the case that
images taken in field conditions suffer from various distortions due
to the position of a module in front of a camera, lens distortions,
blurring due to wind and shocks. Such distortions introduce
complications in automatic image processing. Therefore, a certain
amount of work has been done in the direction of EL image
preprocessing analysis methods \cite{deitsch2018segmentation,
  karimi2019automated, sovetkin2019automatic}.

The preprocessing steps are important for IR images as
well. \citeauthor{salamanca2017detection}
\cite{salamanca2017detection} use the grey-level co-occurrence matrix
to identify the location of the solar panels in IR images of the
operating photovoltaic plants.



\subsection{Image analysis in other fields}

Convolution neural networks are becoming a standard tool in
automated image analysis. We provide here several references that use
similar methods as used in this work.

\citeauthor{masci2012steel} \cite{masci2012steel} developed a CNN
architecture to detect cracks in steel. The proposed architecture is
compared with a method that use hand-crafted feature description.

A neural network architecture has been proposed to identify cracks on
the road, \cite{zhang2016road}. It has been demonstrated that the
method outperforms methods based on SVM and Boosting.

\citeauthor{waldner2019deep} \cite{waldner2019deep} use
encoder-decoder neural networks to extract agriculture field
boundaries from satellite images. \citeauthor{iglovikov2018ternausnet}
\cite{iglovikov2018ternausnet} use \Nunet{} with VGG11 encoder network to
segment satellite images in the Inria Aerial Image Labelling Dataset
\cite{inriadataset}.

Image segmentation is an important topic in medical image analysis as
well. \citeauthor{havaei2015convolutional}
\cite{havaei2015convolutional} performs semantic segmentation of the
brain tumours using MRI imaging. They explore the possibility to combine a
simple CNN in a cascaded fashion. \citeauthor{esteva2017dermatologist}
\cite{esteva2017dermatologist} use deep neural network to classify
different types of skin cancer. \citeauthor{attia2017surgical}
\cite{attia2017surgical} use combination of CNN and Recurrent Neural
Networks (RNN) to identify a surgical tool location in medical
imaging. \citeauthor{kayalibay2017cnn} \cite{kayalibay2017cnn} adapt
the \Nunet{} encoder-decoder style architecture for the 3-dimensional
input signal of the MRI images.

\section{Data}
\label{sec:data}

The data was acquired within the framework of the PEARL-TF project. The
website~\cite{pearltf_eu} contains detailed information about the
project and the involved partners. In this project, the data from
several solar parks with thin-film modules was collected. In addition
to EL images, also performance characteristics of the modules were
measured.

The EL images are taken at predefined conditions (selected fixed applied
current and/or fixed applied voltage). A silicon CCD sensor camera is
used to measure subsequently several parts of the module, with the
images being stitched afterwards. The applied voltage and the applied
current together with the temperature of the module are being recorded.
The I/V characteristics are also measured and the solar cell performance
parameters determined.

\begin{figure}
  \centering
  \includegraphics[angle=90,width=0.95\linewidth]{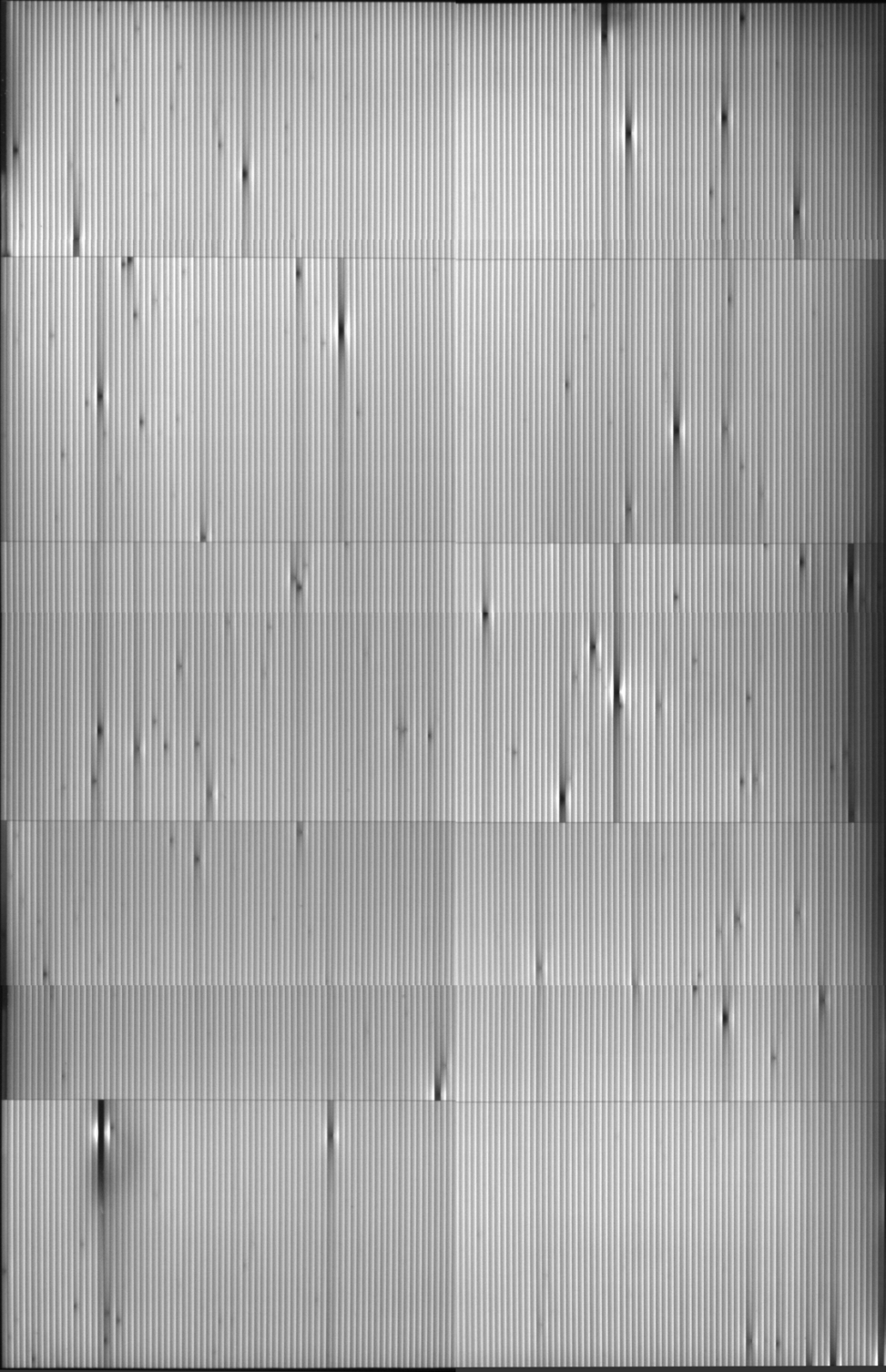}
  \caption{Thin-film module EL image. A module consists of 150 cells
    (positioned horizontally) connected in series. The cells are
    separated by interconnection lines (horizontal dark lines). The
    module consist of several submodules separated by vertical
    isolation lines, which appear dark in the EL image. The EL image is
    stitched (there are 1 horizontal and 3 vertical stitch lines);
    overall intensities of different patches of images are different.
    These intensity differences are attributed to metastable changes
    during the measurement.}
  \label{fig:module}
\end{figure}

\linelabel{shorten data description}The database contains 6000 EL
images of the co-evaporated Copper Indium Gallium Diselenide (CIGS)
modules from the same manufacturer. Every image is supplied with a
measured performance data. A typical EL image of a thin-film CIGS
module from our database is depicted in Figure~\ref{fig:module}. The
module consists of 150 connected cells in series (in
Figure~\ref{fig:module} the cells are recognized as horizontal
stripes). The cells are separated by interconnection
lines 
(horizontal dark lines in Figure~\ref{fig:module}). In addition, the
module is separated in 5 parallel sub-modules by vertical isolation
lines (dark vertical lines).

As mentioned before, every EL image consists of several stitched
images.  Different stitched parts of the image have different overall
intensities (see Figure~\ref{fig:module}). This is attributed to the
metastable behaviour of CIGS solar cells, where the electrical
properties of the cell can change during the measurement.

In order to obtain\linelabel{manual labelling} a labelled dataset we
segment images manually. This work is done using two different image
editor programs. Firstly, we use the GNU\footnote{GNU is a recursive
  acronym for GNU's Not Unix!}  Image Manipulation Program
(GIMP)~\cite{gimp} to create binary masks of various defects
locations, where the defect pixels are manually marked using a drawing
pad and a digital stylus. Alternatively, we use the
ThinFia~\cite{thinfia} program that is designed to identify defects in
EL images by introducing a grid-mesh. \linelabel{thinfia and gimp}The ThinFia program was
developed within the PEARL-TF project. A general image processing
program such as GIMP requires more time to segment an image, comparing
to the ThinFia, however, smaller defects are segmented more accurately
in GIMP. The annotations for droplets and shunts are each performed by
a single person, and thus the inter-annotator agreement has not been
considered.\linelabel{inter-annotator}

\begin{figure}
  \centering
  \begin{minipage}{.48\linewidth}
    \includegraphics[width=.95\linewidth,height=.65\linewidth]{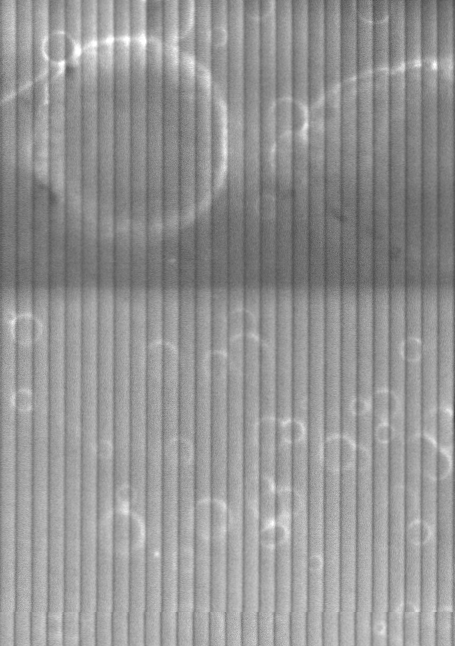}
  \end{minipage}
  \begin{minipage}{.48\linewidth}
    \includegraphics[width=.95\linewidth,height=.65\linewidth]{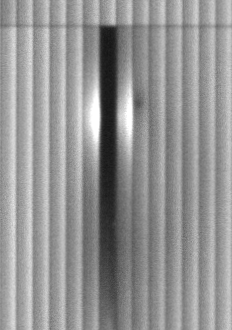}
  \end{minipage}
  \caption{Example of droplets (left) and a shunted area (right). Here
    the cells are shown vertically.}
  \label{fig:shunts and droplets}
\end{figure}

In the image database we focus on the segmentation of
``shunts''. Shunts are characterized by a more conductive connection
between the front and back electrodes than the normal solar cell
structure (i.e.\ the solar cell structure is damaged or
missing). There are many causes for shunts. Commonly shunts originate
from debris of the copper evaporation source or pinholes in the CIGS
absorber \cite{misic2015debris, misic2017thesisoriginofshunts}.
Figure~\ref{fig:shunts and droplets} (right) depicts a shunt
defect. Shunts generally appear as dark areas with a gradient in
intensity away from the actual defect location. The dark area is
confined to the area of one cell. Severe shunts may also completely
darken a cell stripe, in which case often the neighboring cells
exhibit bright areas in the vicinity of the shunt
\cite{tran2011characterization}. Shunts are generally relevant to the
solar module performance, in particular under low light conditions
\cite{weber2011electroluminescence}.



In addition to shunts we noticed the CIGS modules often exhibit
``droplets'' in the EL images. Figure~\ref{fig:shunts and droplets}
(left) shows a detail of droplets. The appearance of droplets
resembles water stains and thus we speculate these structures
originate from the chemical bath deposition. At this point it is
unknown what the impact of droplets is on the module performance,
however, the bright appearance imply a local change in quantum
efficiency according to the reciprocity relations between luminescence
and quantum efficiency\linelabel{typo: page 3 efficiency}
\cite{Rau2007Reciprocity}.

In total\linelabel{total images}, we have about 6000 unlabelled, 142
labelled module images with shunts, and 14 labelled module images with
droplets. The manual segmentation of droplets in an image is
particular laborious, hence only few images are available.

\linelabel{training and testing datasets} All labelled images are
split randomly onto a training and a testing datasets. The training
dataset consists 106 labelled shunts and 8 labelled droplets
images. The testing dataset contains 24 labelled shunt and 3 droplets
images. In addition we evaluate the final model using the remaining 12
images with shunts and 3 images with droplets.

The labelled dataset is available online, \cite{data}.

\section{Methods}
\label{sec:methods}

In this section we review methods that we use to build segmentation
models. The {\it encoder-decoder} deep neural network architectures,
\cite{goodfellow2016deep}, are commonly used in the semantic image
segmentation problems,
\cite{kayalibay2017cnn,attia2017surgical,iglovikov2018ternausnet,waldner2019deep}. We
build multiple models by combining different encoder and decoder parts
in the neural network architecture.

In order to populate our training data set we apply various
transformations (data augmentation) to the original EL images.

After several models are trained, we compute segmentation on the test
images and compute several performance metrics. It is important to
establish a baseline for the evaluated metrics, as manual labelling
has a significant human bias and differs for different types of
defects.

The final model is obtained by means of using the multi-objective
optimization technique. 

It should be noted here, that the encoder-decoder models used here
solve a general image segmentation problem, and hence can be applied
to arbitrary images, as long as there exist a training dataset of
sufficient size. For instance, these methods can be applied to
different defects, different types of images (e.g.\ IR, PL, UVFL) or
PV technologies (e.g.\ crystalline modules).

Our current approach is only restricted by the size of the input image
patch and information contained in it. Therefore, defects that are
larger than the input image patch can be problematic to identify.

\subsection{Model architectures}

For our segmentation models we utilize the encoder-decoder neural
network architecture. These networks consist of two parts: the
contraction part (or encoder) and the symmetric expansion part (or
decoder). The encoder compresses information content of an arbitrarily
high-dimensional image into a feature vector. The decoder gradually
upscales the encoded features back to the original
resolution.

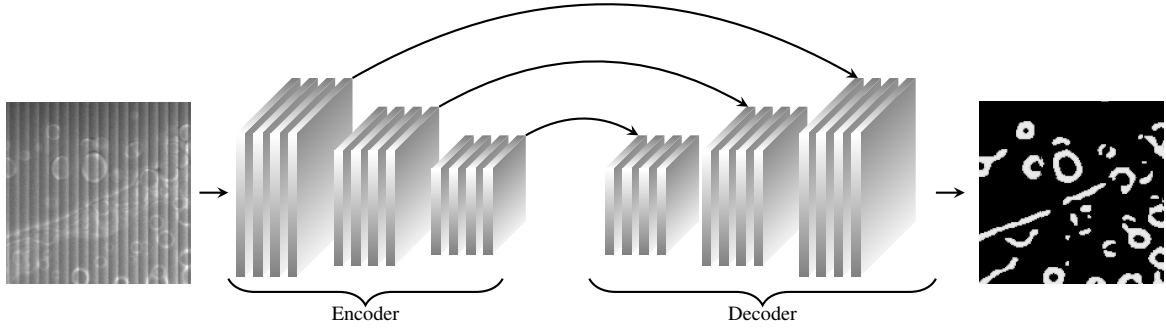
\begin{figure*}
  \centering
  \begin{tikzpicture}[scale = 0.38, transform shape]
    \tikzset{pics/layer/.style n args={2}{
        code = { %
          \begin{scope}[canvas is zy plane at x=0]
            \path[very thin,shading angle=45] (0,0) rectangle (#1,#1);
          \end{scope}
          \begin{scope}[canvas is xy plane at z=#1]
            \path[very thin,shading angle=180] (0,0) rectangle (-#2,#1);
          \end{scope}
          \begin{scope}[canvas is zx plane at y=#1]
            \path[very thin,shading angle=0] (0,0) rectangle (#1,-#2);
          \end{scope}
        }
      }
    }
    \def\d{0.3}
    \foreach \x in {0,...,3}
    \pic at (2*\d*\x,0) {layer={5}{\d}};
    \foreach \x in {0,...,3}
    \pic at (2*\d*\x + 2*5*\d,0) {layer={4}{\d}};
    \foreach \x in {0,...,3}
    \pic at (2*\d*\x + 2*10*\d,0) {layer={3}{\d}};
    \foreach \x in {0,...,3}
    \pic at (2*\d*\x + 12,0) {layer={3}{\d}};
    \foreach \x in {0,...,3}
    \pic at (2*\d*\x + 2.5*5*\d + 12,0) {layer={4}{\d}};
    \foreach \x in {0,...,3}
    \pic at (2*\d*\x + 2.5*10*\d + 12,0) {layer={5}{\d}};
    \draw [thick,decorate,decoration={mirror,brace,amplitude=10pt}]
    (-2.5,-2) -- (7,-2) node
    [black, midway, anchor=north, yshift=-24pt] {\huge Encoder};
    \draw [thick,decorate,decoration={mirror,brace,amplitude=10pt}]
    (10,-2) -- (22,-2) node
    [black, midway, anchor=north, yshift=-24pt] {\huge Decoder};
    \node at (-7,1) {\includegraphics[width=.35\textwidth]{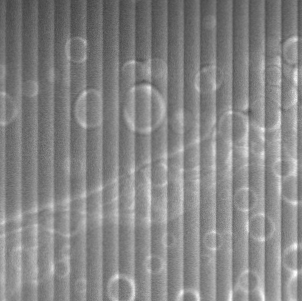}};
    \node at (26.7,1) {\includegraphics[width=.35\textwidth]{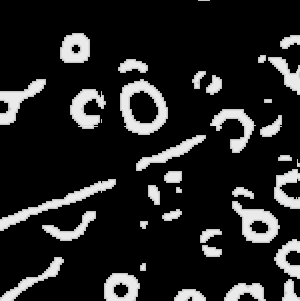}};
    \draw [->,>=stealth,thick] (-3.5,1) -- (-2.5,1);
    \draw [->,>=stealth,thick] (22,1) -- (23,1);
    \path [->,>=stealth,thick,bend left] (7.8,3) edge (11.7,3);
    \path [->,>=stealth,thick,bend left] (4.8,4) edge (15.5,4);
    \path [->,>=stealth,thick,bend left] (1.8,5) edge (19.3,5);
  \end{tikzpicture}
  \caption{Schematic representation of an encoder-decoder
    architecture. The left-hand side image is an input image patch
    that is passed to a series of the computational layers. The
    right-hand side image is an output binary image. The arrows are
    skipping connection layers, where input is being copied directly
    from encoder to a decoder.}
  \label{fig:encoder-decoder}
\end{figure*}

Figure~\ref{fig:encoder-decoder} schematically demonstrates a typical
structure of the encoder-decoder architecture. Each block represents
an output of the convolutional layer, with the data flow going from
left to right. The arrows represent the skipping links, where the
input for the layer is copied from the encoder to the decoder
parts. Different networks may have different number of layers,
skipping connections and different activation functions.

We take several popular networks and use parts of their architecture
as an encoder or as a decoder. Combinations of encoders and decoders
from different networks provide us with different segmentation
networks. Furthermore, there are several tuning parameters available
that allow us to generate even more segmentation models.

For the encoder part of our networks we use \Nmobilenet{}
\cite{howard2017mobilenets}, \Nresnet{} \cite{he2016deep}, \Nvggnet{}
\cite{simonyan2014very} and \Nunet{} \cite{ronneberger2015u}. For the
decoder part of our network we use \Nunet{} \cite{ronneberger2015u},
\Nfcnnet{} \cite{long2015fully}, \Npspnet{} \cite{zhao2017pyramid} and
\Nsegnet{} \cite{badrinarayanan2017segnet}. We use network
implementations available in \cite{keras_models}.

In order to be able to combine different encoder and decoder
architectures we organize all encoder outputs to have 5 feature
tensors. Different models have slightly different permissible input
image sizes. The \Nresnet{} models require the input image
dimension to be divisible by 32, the \Npspnet{} models: divisible by
192. The \Nmobilenet{} models have the fixed input image size of
$224 \times 224$ pixels.

Table~\ref{table:encoder decoder combination} summarizes the
combinations of the encoder and decoder models. Some encoders and
decoders have a tuning parameter. \linelabel{tuning parameters} The
\Nresnet{}, \Nvggnet{}, \Nunet{} have the size of the input image as
an input parameter. We select between either $256\times 256$ and
$512\times 512$. The \Nsegnet{}'s parameter is the number of
upsampling routines, see \cite{badrinarayanan2017segnet}. The
\Nfcnnet{}'s parameter has a single parameter: number of features,
that can be equal $8$ or $32$. Combinations of different encoders and
decoders with the choice of several tunable parameters provide us with
a choice of 34 trainable models\linelabel{tuning parameters result in
  models}, i.e.\ we treat a single model with tuning parameters as
several independent models corresponding to different choices of
tuning parameters. Some of the \Nfcnnet{} models with 32 layers are
discarded as they require too much memory.

\begin{table}
  \centering
  \caption{Various combination of encoder and decoder networks}
  \label{table:encoder decoder combination}
  \begin{minipage}[t]{.5\linewidth}
    \begin{tabular}{c|c}
      Encoder & Decoder \\
      \hline
      \Nmobilenet{} & \multirow{3}{*}{\Npspnet{}} \\
      \Nresnet{} &  \\
      \Nvggnet{} &  \\
      \hline
      \Nmobilenet{} & \multirow{4}{*}{\Nsegnet{}} \\
      \Nresnet{} &  \\
      \Nunet{} &  \\
      \Nvggnet{} &
    \end{tabular}
  \end{minipage}%
  \begin{minipage}[t]{.5\linewidth}
    \begin{tabular}{c|c}
      Encoder & Decoder \\
      \hline
      \Nmobilenet{} & \multirow{4}{*}{\Nunet{}} \\
      \Nresnet{} &  \\
      \Nvggnet{} &  \\
      \Nunet{} &  \\
      \hline
      \Nmobilenet{} & \multirow{4}{*}{\Nfcnnet{}} \\
      \Nresnet{} &  \\
      \Nvggnet{} &  \\
      \Nunet{} &
    \end{tabular}
  \end{minipage}
\end{table}


All models work with input image patches that are smaller than a
typical EL image. The size of the input image regulates the
information that is used for defect forecast. As shunts defects and
droplets have a local nature, we select a square image patch size with
an edge length varying between 15 and 30 times the cell width.

\subsection{Data augmentation}
We apply data augmentation to the training data set. The data
augmentation is performed in order to increase the number of images to
train our models with. This is particularly important for recognizing
droplets as we have few labelled images to train with.

For the data augmentation, it is important to note that convolutional
neural networks are able to incorporate spatial information in its
model, however, they are not equivariant to scale and rotation
transformations. Thus, the augmentation of data must entail scaling
and rotation operations in order to boost the ability of the network
to generalize, \cite{goodfellow2016deep}.

We organized data augmentation procedures in a pipeline of data
augmentation generators. A generator accepts an image as its input and
outputs a list of augmented images. Every image in the output is
``piped'' to the following generator in the pipeline. Thus, a single
input image generates a series of augmented output images.

Every data augmentation generator has a set of tunable
parameters. These parameters regulate the properties of the generated
output images, such as size and transformation severity.

First augmentation transformation we perform are done on the original
pair of EL and the corresponding label images. An image is re-sized
preserving the aspect ratio, with the scaling factors ranging in the
interval $[0.7,1.3]$.  This scaling operation allows to populate the
training dataset with examples of modules of slightly different cell
sizes. After scaling, we perform mirroring of the image with respect to
the vertical and horizontal axes, and apply the following contrast
transformation to the EL image:
\begin{gather}
  \big( (1-\alpha)M - (1+\alpha)m \big) \frac{I - m}{M-m} + m(1+\alpha),
\end{gather}
where $M$ and $m$ are the maximum and the minimum values of the image,
$\alpha$ is selected in a range of values $[-0.4,0.4]$.

Our segmentation models use a particular dimension of the input
image. For example, as mentioned before \Nmobilenet{} accepts only
images with the width and height of 224 pixels, and the dimensions of
the \Nresnet{} should be divisible by 32. For this reason the final
generator in the pipeline generates the subimages of the required
dimensions. The subimages are generated using a sliding window of the
selected size. The window shift is chosen to be 50 pixels, thus, we
obtain overlapping subimages.

The \Nresnet{} networks input image size is chosen to be equal of
width and height $256$ and $512$ pixels, the \Nmobilenet{} network ---
224 pixel and the \Npspnet{} network --- 192 pixels.\linelabel{input}

The defects are distributed in a non-uniform way. Therefore, to have a
representative sample, we select only those subimages that contain at
least 200 pixels corresponding to a defect\linelabel{augment filter}. All the training images
are shuffled, so that, a particular order of the generator output does
not play any role.

\subsection{Training}

We use the transfer learning technique \cite{goodfellow2016deep},
i.e.\ we use the pre-trained model on an unrelated dataset of images,
and fine-tune it to adapt to the new task. We use weights that were
pre-trained on the ImageNet dataset, \cite{imagenet_cvpr09}. The
weights for various networks are accessible in
\cite{fchollet_weights}.

In our training we use the dropout regularization technique, with the
dropout rate chosen $0.2$; as well as the batch normalization
technique. The latter method allows to normalize the inputs of every
layer in the network, that helps reduce effect of the so-called
covariance shift, \cite{ioffe2015batch}.

For the loss function we use the categorical cross-entropy function,
\cite{goodfellow2016deep}\linelabel{loss function}. We use the
AdaDelta optimizer \cite{zeiler2012adadelta}, as it requires no manual
setting of a learning rate; it is robust to large gradients, noise
architecture choice and insensitive to hyperparameters. The number of
epochs is chosen to equal 100 with 512 gradient steps in each
epoch. The training process stabilizes by this time, with no
overfitting being observed in accuracy metrics evaluated on a testing
dataset.

\subsection{Accuracy metrics}

To compare the trained segmentation models we compute performance
metrics on pairs consisting of a segmented image and the corresponding
ground truth mask.

The first performance metric is a common Jaccard index,
\cite{van1979information}. Given two binary masks $A$ and $B$, the
Jaccard index is defined by
\(
  J(A,B) \coloneqq \frac{|A\cap B|}{|A \cup B|},
\)
where $|A|$ denotes the number of non-zero elements in the binary mask
$A$. The Jaccard index attains values in the interval $[0,1]$, where
$0$ corresponds to the case when the binary masks $A$ and $B$ do not
have common values.

We evaluate an index on a set of test images, thus obtaining a sample
of index values for each model. \linelabel{our empirical finding
  remove} The Jaccard index can be used to discover very
bad-performing models, however, it cannot identify a significant
difference between models.\linelabel{visuallydifferentmodels} We
conjecture that this happens because the Jaccard index reflects not
only errors in defect locations, but also in its shape.

To address this drawback we introduce a second set of performance
metrics, that ignores the shape information and focuses on the
accuracy of the defects identification. We assume here that a single
connected component in a segmented image constitutes a defect.

To this end, let $A$ and $B$ be two binary masks, and $K(A)$ is a set
of connected components in $A$. Define the {\it component instance
  function}
\begin{gather}
  \label{eq:1}
  I(A,B) \coloneqq
  \frac{
    \big| x \in K(B) \,|\, \exists y \in K(A \cap B):
    y \cap x \neq \emptyset \big|
  }{|K(B)|}.
\end{gather}
Note that the component instance function attains values in the
interval $[0,1]$, as the nominator is always less or equal than
denominator. The connected components are computed using the labelling
algorithm, \cite{wu2005optimizing}.

The instance function is not symmetric, and if $A$ is a segmentation
output and $B$ is the ground truth, then $I(B,A)$ can be interpreted
as a precision index and $I(A,B)$ can be interpreted as a recall
index.

The precision and recall are typical metrics computed in
classification problems. In our case they can be interpreted in the
following way. \linelabel{true location explanation}The precision is
the proportion of correctly identified defects among all locations
identified by a model, while the recall is the fraction of all defects
that were identified by a model.

The component instance function formalises the meaning of a single
defect by using the notion of a connected component. The defect is
identified correctly, if the component on the ground truth image and
the segmented image have a non-zero intersection.

\subsection{Metrics baseline identification}

A typical accuracy metric (or index) attains values in the interval
$[0,1]$, where $0$ corresponds to a bad segmentation and $1$ to a
perfect segmentation. In practice, however, the value of $1$ cannot be
achieved. The reason for that in our case is the lack of clearly
defined ground truth segmentation.

This happens due to several reasons. Firstly, when a defect expresses
itself with a smooth gradient change in the image pixel value
intensity, the beginning of the defect can be
ambiguous\linelabel{ambiguous labelling}. For example, this can be
observed in a shunted area, see Figure~\ref{fig:shunts and droplets}
(right). A shunt can be a microscopic defect, but on an EL image it
induces a much larger darkened region with a smooth gradient
change. Secondly, there can be too many defects to label them manually
with high accuracy. For example, Figure~\ref{fig:shunts and droplets}
(left) shows a part of a module with so-called droplets.

\linelabel{establishing a baseline} Therefore, to identify the values
of the metrics that can be considered as good, we propose the
following procedure. A set of images are segmented manually twice by
the same person. A sufficient amount of time (more than 1
week) is taken between the first and second manual
segmentation.\linelabel{double segmentation baseline} The resulting
segmentation images are evaluated using the performance metrics, to
obtain the baseline values for the metrics.

\subsection{Model selection}

Combinations of different encoder and decoder networks with different
tuning parameters yield a set of segmentation models. In this section
we address the question of how to select the best model from this set,
and discuss related problems.

Previously, we defined several metrics that are capable of measuring
the quality of the segmentation given a manually segmented image. A
two-stage procedure as described above allows to identify the baseline
values for each metric, which allows to normalize the values and avoid
the human bias incorporated in the raw metrics.

However, it is not possible to select the best model based on a single
metric value. Firstly, from a sample of test images a single metric
cannot significantly distinguish between several well performing
models. At the same time, a visual inspection of the evaluations of
this model indicate that models differ. Secondly, different metrics
incorporate different information in its values. For example, the
precision metric $I(B,A)$ measures how accurate are the identified
defects by the model, whereas the recall metric $I(A,B)$ measures the
total percentage of all identified defects.

Therefore, in order to address these problems we propose the following
heuristic two-stage procedure. In the first step we select a set of
models using a multi-objective optimization similar to the one used by
\citeauthor{waldner2019deep}, \cite{waldner2019deep}. We use our
precision and recall indices as the targets in the multi-objective
optimization. In the second step, from a set of chosen models we
select the final model using the Jaccard index. \linelabel{modelselectionfrompareto} 

\section{Results}
\label{sec:results}

To compute evaluation of a full sized module image we compute
evaluation on a set of overlapping subimages, where the size of each
subimage equals to the selected model input image size. Note that this
introduces a limit on the size of the structures that can be detected
as the model cannot properly detect structures larger than the
subimage size. The resulting segmented image patches are combined with
the overlapping borders being removed.

\linelabel{overlapping} We overlap images by selecting a shift of 10
pixels less than what would be needed for producing subimages without
an overlap. The width of the removed subimage borders equals to 5
pixels. The areas on the borders of the original image are not
overlapped, and hence not removed.

The purpose of overlapping is to mitigate an incorrect segmentation
close to the edges of the image patches. This problem is observed
frequently in the \Nfcnnet{} networks.

After the image augmentation pipeline we have about 15000 training
images for shunts and 5000 training images for
droplets.\linelabel{augment number images}

Below we discuss the model selection for shunts and droplets, we
demonstrate examples of shunts and droplets segmentation for the
selected models, and provide details on the implementation and time
required for the segmentation.

\subsection{Model selection}

\begin{figure}
  \centering
  \includegraphics[width=0.7\linewidth,height=0.55\linewidth]{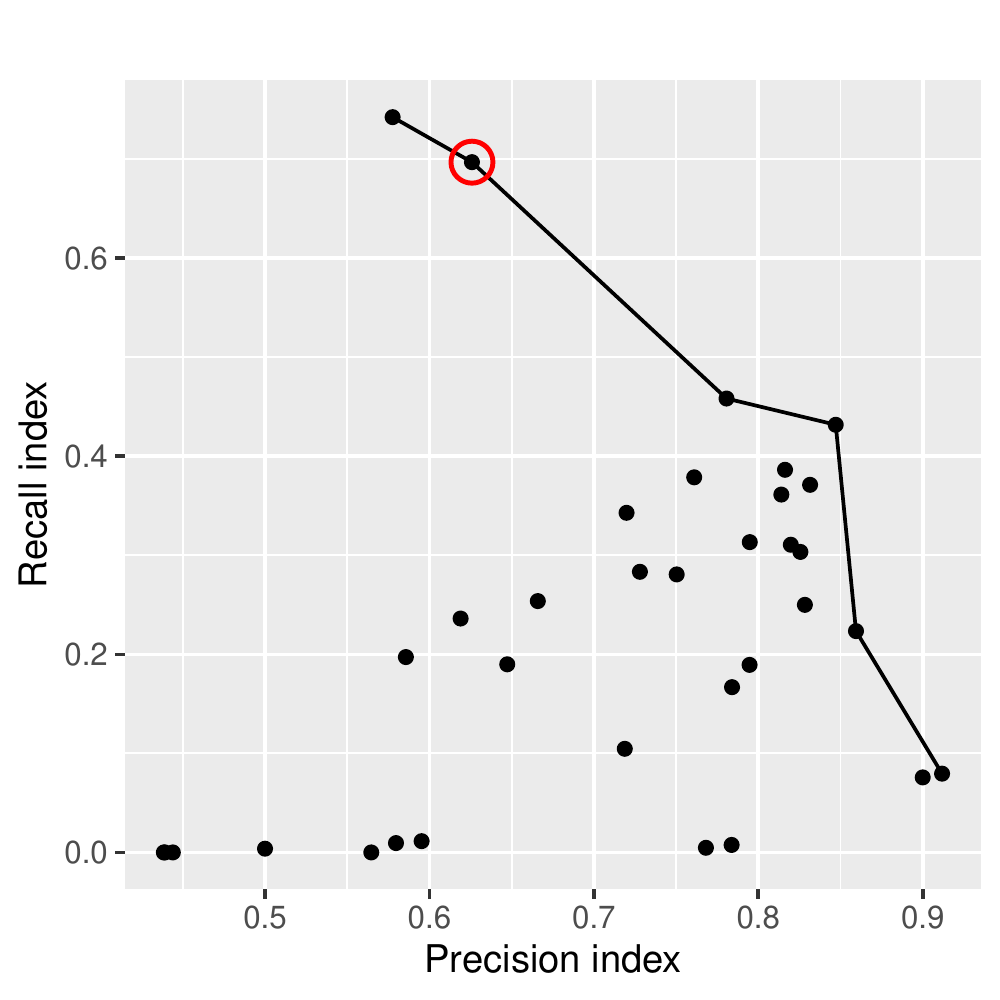}
  \includegraphics[width=0.7\linewidth,height=0.55\linewidth]{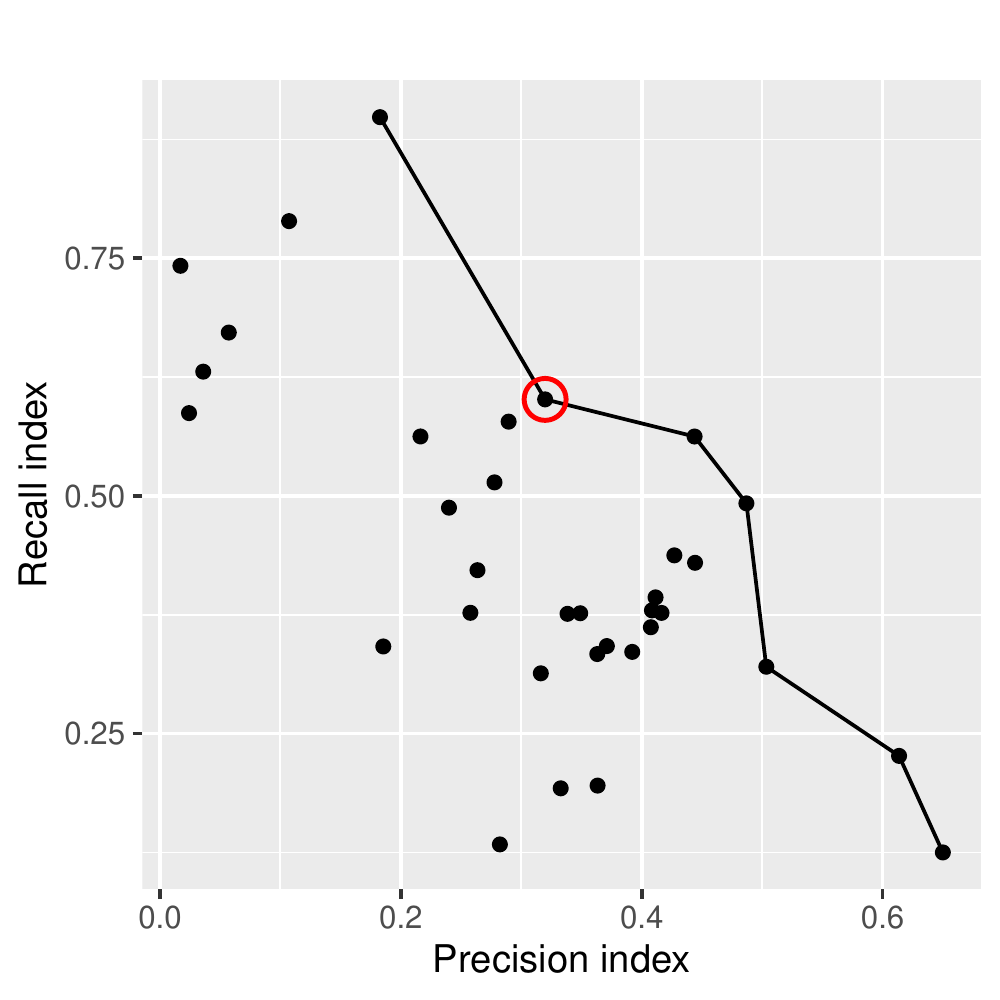}
  \caption{Pareto frontier for the droplets (top) and shunts (bottom)
    models}
  \label{fig:frontier droplets}
\end{figure}

Figure~\ref{fig:frontier droplets} shows\linelabel{figure 4 and 5
  clarifications} the estimated median of the precision and recall
indices for each of the model for droplets (top figure) and shunts
(bottom figure). The model evaluations are performed on a testing
dataset that was not used during the training
process\linelabel{evaluation on testing dataset}.

The black line is the Pareto frontier of precision/recall
multi-objective optimization for the droplets and shunts models. The
Pareto frontier allows to select 6 models for droplets and 7 models
for shunts. The red circle indicate the model on the frontier that
maximizes the Jaccard index\linelabel{modelselectionfrompareto2}. For
the shunts the best model is \Nmobilenet{}-encoder and
\Nfcnnet{}-decoder, and for the droplets the best model is
\Nvggnet{}-encoder and \Nunet{}-decoder.

Note that the estimated baseline for the precision and recall indices
are equal 0.8 for both shunts and droplets, the Jaccard index baseline
equals to 0.24 for shunts and 0.34 for droplets.\linelabel{baseline
  estimated}

\linelabel{tuning parameters 2} All selected models have an input image
of dimensions $256 \times 256$ pixels (one of the tuning parameter for
\Nfcnnet{} and \Nunet{}), and the kernel size of the \Nfcnnet{}
network equals to 8.

\linelabel{final evaluation} We perform a final evaluation on a set of
images disjoint from training and testing images.
Table~\ref{table:final evaluation} summarised the median value of the
precision, recall and Jaccard indices evaluated on 24 shunts and 3
droplets images.

\begin{table}
  \centering
  \caption{Final evaluation of the selected models}
  \label{table:final evaluation}
  \begin{tabular}{l|c|c|c}
    \backslashbox{Feature}{Index} & Precision & Recall & Jaccard \\
    \hline
    Shunts & 0.35 & 0.55 & 0.19 \\
    \hline
    Droplets & 0.61 & 0.68 & 0.27 \\
  \end{tabular}
\end{table}

\begin{figure}
  \centering
    \includegraphics[width=0.9\linewidth,height=0.27\linewidth]{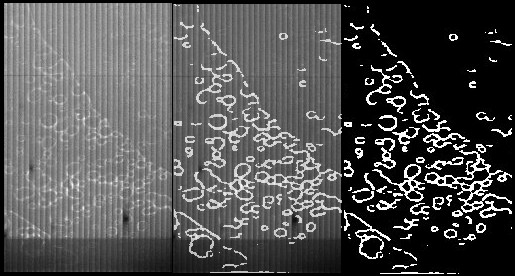}
    \includegraphics[width=0.9\linewidth,height=0.27\linewidth]{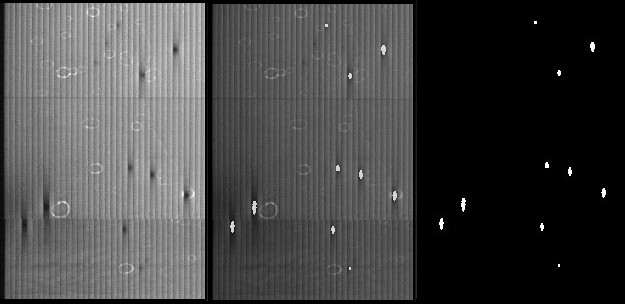}
    \caption{An example of droplets (top) and shunts (bottom)
      segmentation}
    \label{fig:segmentation shunts}
\end{figure}

In Figure~\ref{fig:segmentation shunts} we show examples of the
segmentation of droplets and shunts on two thin-film modules using the
selected models. On the left side of the figures is shown the original
image. On the right side is the binary mask of the segmentation. In
the centre is shown an original image overlayed with the binary mask.

\subsection{Segmentation examples and heat-maps}

\begin{figure}
    \begin{minipage}{\linewidth}
      \centering
      \includegraphics[width=.5\linewidth,height=0.9\linewidth,angle=270]{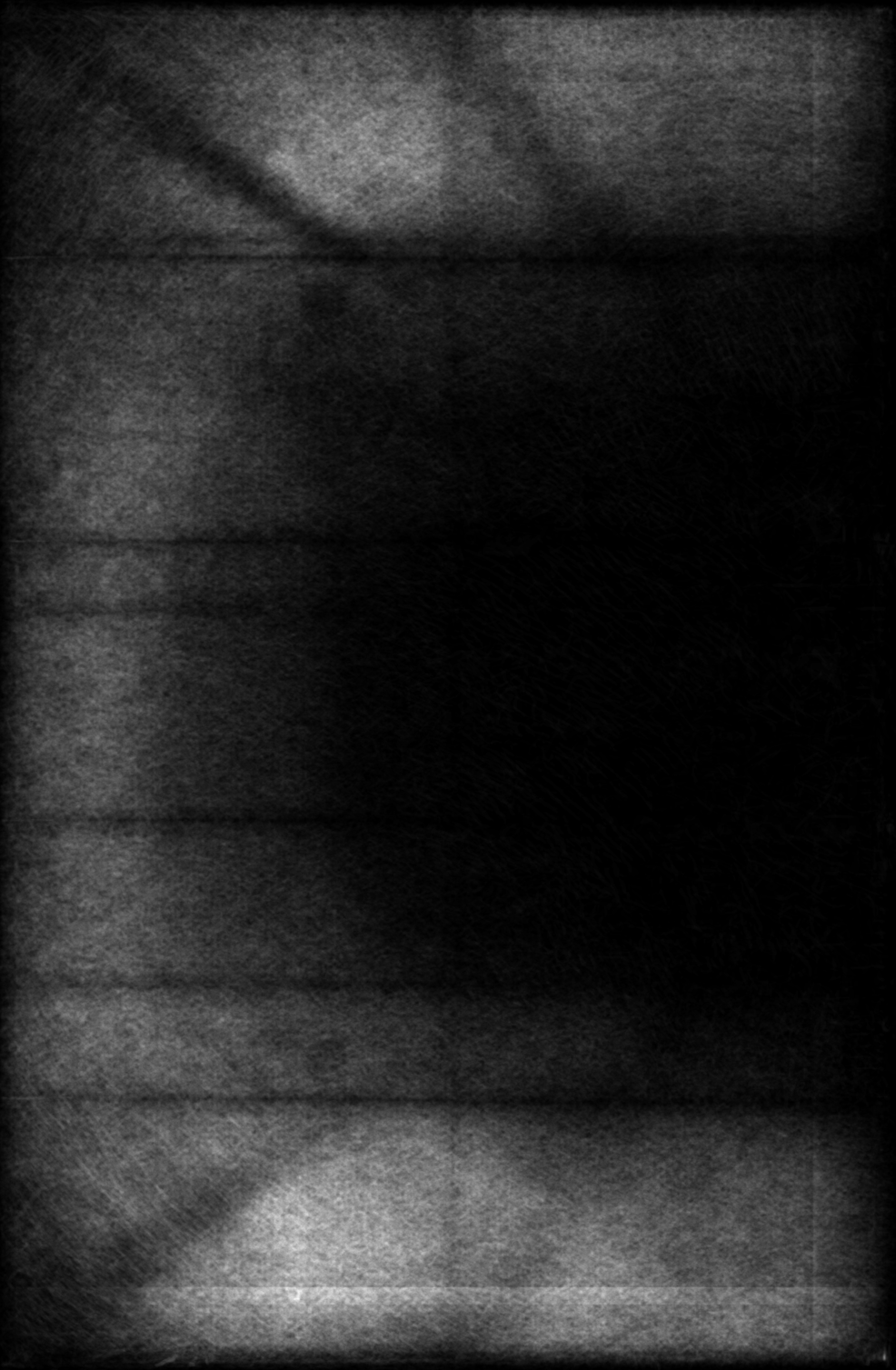}
      \includegraphics[width=.7\linewidth]{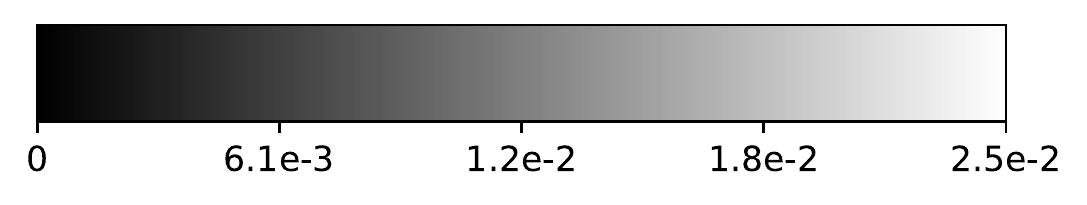}
    \end{minipage}
    \begin{minipage}{\linewidth}
      \centering
      \includegraphics[width=.5\linewidth,height=0.9\linewidth,angle=270]{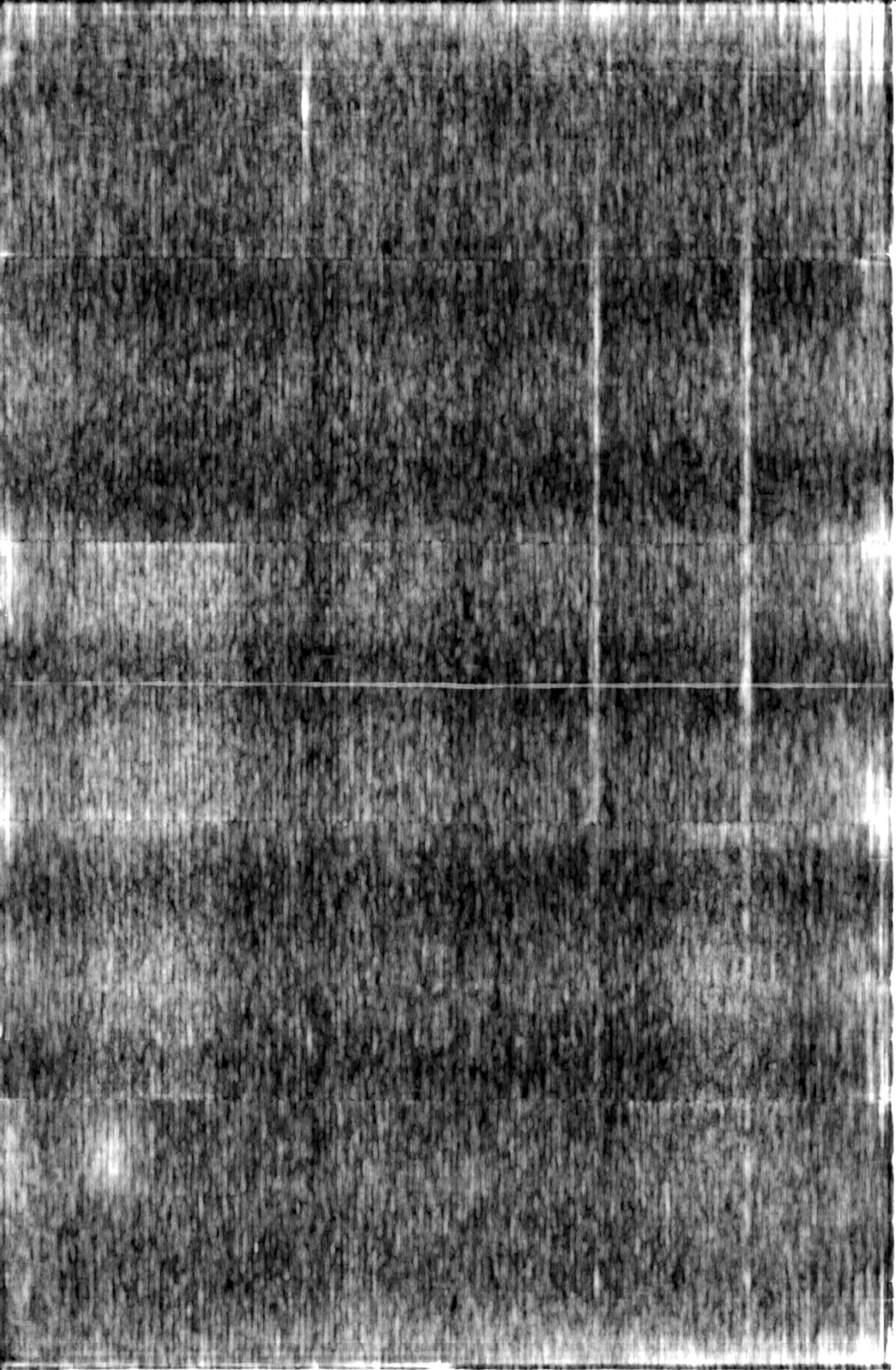}
      \includegraphics[width=.7\linewidth]{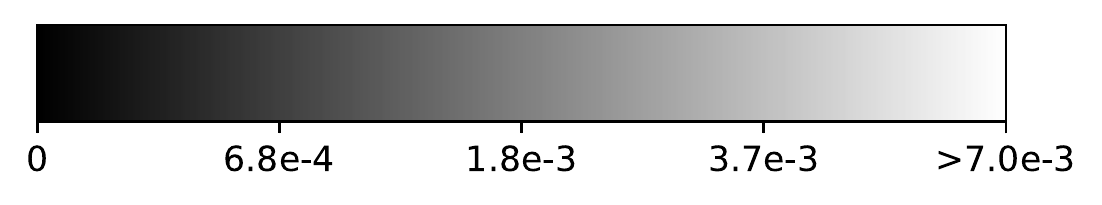}
    \end{minipage}
    \caption{Heat map of droplets (top) and shunts (bottom)
      locations. The intensity scale indicates the probablity a pixel
      is marked as a droplet or a shunt.}
    \label{fig:heatmap droplets}
\end{figure}

We apply the segmentation model for droplets and shunts for each EL
image in our database. By evaluating an average of the computed binary
segmented image, we obtain so-called heat maps.
Figure~\ref{fig:heatmap droplets} depicts heat maps based on 6000 EL
images for droplet (top figure) and shunts (bottom figure)
locations\linelabel{CIGS subset}. The brighter areas correspond to
locations where defects have a higher probability of occurrence. The
heat maps are given with a scale that maps pixel value in heat map
images to a probability\linelabel{heatmap probability} of shunts or
droplets in that pixel. Note, that the scale is logarithmic in
Figure~\ref{fig:heatmap droplets} (bottom), with the $1\%$ of the
brightest observations are shown as the white color.

The droplets in Figure~\ref{fig:heatmap droplets} (top) expresses a
clear structure where droplets are distributed along a broad arc along
three edges of the module. In the brighter areas of the arc, the
probability that a pixel is marked as a droplet is about 2.5\% (148
times in 6000 images\linelabel{number of images in heatmap}). The
droplets do not occur in the center of the modules.  Furthermore,
there are several dark lines where fewer droplets are detected. The
vertical and horizontal dark lines appear to be interference of
vertical isolation lines in the modules as well as stitching lines in
the image. However, the diagonal lines do not correspond to any
obvious structure in the images that may interfere with droplet
detection. We infer the diagonal lines have a physical origin.

The heat map of shunts is shown in Figure~\ref{fig:heatmap droplets}
(bottom).  There is a great number of features to be seen, such as a
clear banded structure, high concentrations at certain edges and
locations, such as at the bottom edge where at the isolation lines
high concentrations of shunts are detected. Two cell stripes in the
bottom half of the module are more often shunted. The stitching lines
do not show up and thus do not seem to interfere with shunt
detection. The isolation lines are associated with more detected
shunts. However, only parts of the isolation lines exhibit a larger
concentration of shunts and not all isolation lines are equally
affected. For this reason we believe the higher shunt probability
around the isolation lines is no artifact. There is a bright vertical
line in the center. This line does not correspond to the position of
an isolation line or stitching line. Note that in
Figure~\ref{fig:segmentation shunts} a slightly darker vertical line
is visible at the same position. However, in this example no shunts
are detected along this line. In other EL images this darker line is
not present (e.g.\ in Figure~\ref{fig:module}). The origin of this
line is unclear.

We would like to note that many features in Figure~\ref{fig:heatmap
  droplets} are rather subtle, and that these features only become
visible when an average of a large number of images is
computed. Furthermore, some of these features are quite certainly
performance and reliability relevant (e.g.\ positions where shunts are
likely to occur). We thus believe the extraction of such features can
give manufacturers a better insight in their production process and
thus contribute to process optimization and quality control.

\subsection{Correlation to performance data}

\linelabel{correlation to performance} Using the computed segmentation
of EL images it is possible to correlate difference characteristics of
the discovered defects in EL images and the performance of the
modules.

Shunts originate from holes in the CIGS absorber,
\cite{misic2015debris, misic2017thesisoriginofshunts}, and primarely
affect the low-light performance and the shunt resistance of PV
modules \cite{weber2011electroluminescence}. A lower shunt resistance
is expected to affect the fill factor and the slope of the IV
characteristics around short circuit conditions. Note that it is a
common practise in industry to count shunts in order to asses module
quality. 

\begin{figure}
  \centering
  \includegraphics[width=0.75\linewidth]{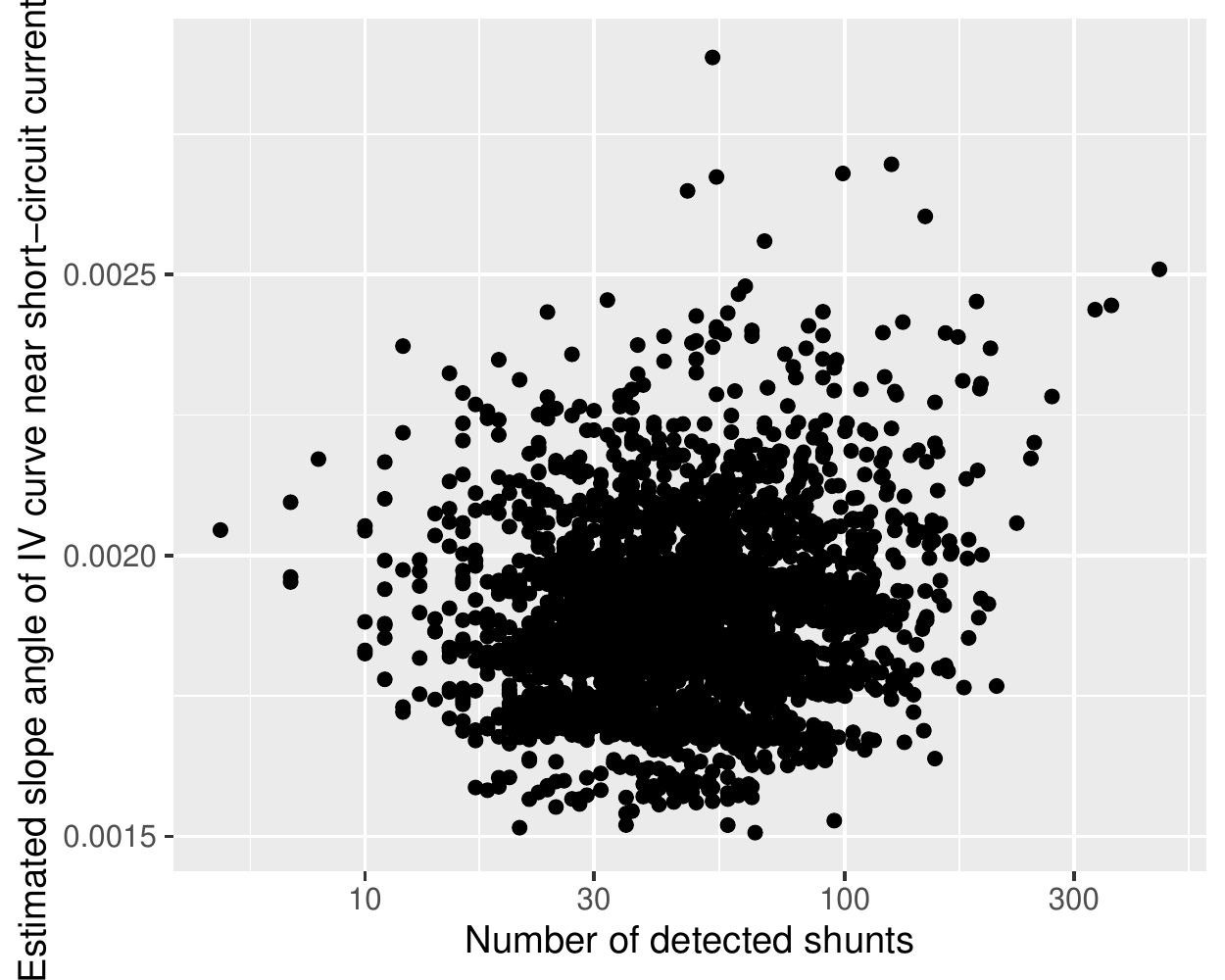}
  \caption{Slope of the IV curve near the short circuit current point
    versus number of shunts detected in the corresponding EL image.}
  \label{fig: slope no components}
\end{figure}

Figure~\ref{fig: slope no components} shows a correlative plot between
the slope of the IV curve near the short-circuit point and the number
of the discovered shunts. The figure reveals no significant
correlations between the variables. Unfortunately, we do not have
low-light performance data for the correlation.

Our study indicated that there are no significant correlations between
size and amount of defects and the module performance data. Detailed
discussion on the correlations between the discovered defects and
performance of the modules is beyond the discussed topic of this
paper.






\section{Conclusions and outlook}
\label{sec:conclusions}

In this paper, we applied the encoder-decoder deep neural networks in
order to perform semantic segmentation of EL images of thin-film
modules. The framework is general and applicable to other types of
defects, PV images (e.g.\ thermography), as well as PV technologies.

We demonstrated the use of encoder-decoder deep neural networks to
detect shunt-type defects and so-called droplets in thin-film CIGS
solar cells. Several models are tested using various combinations of
encoder-decoder layers. A method is proposed to select the best model
based on the collection of metrics that evaluate different accuracy
characteristics.

We show exemplary results for our selected best models of shunts and
droplets. Furthermore, we analyzed a database with 6000 images of CIGS
modules, all of one module type and one manufacturer. We show heat
maps depicting the probability of a shunt or droplet occurring at a
certain location in the solar module. The results show that the
systematic segmentation of a large volume of images can reveal subtle
features which cannot be inferred from studying individual
images. Thus, we argue this type of segmentation models may aid
process optimization and quality control by manufacturers.

Image segmentation methods is an active field of research, and there
are several ways our approach can be extended. Firstly, the ensemble
learning technique can be used to combine several segmentation
models together with an aim to improve the model accuracy. Secondly,
one can use the Generative Adversarial type of Neural networks (GAN)
that can learn the defining features of the sample of images. The
segmentation models can be then built on top of the constructed
network.

Lastly, we remark that the code is available upon request, and the
labelled dataset is available online, see \cite{data}.\linelabel{code
  available and training data}

\section*{Acknowledgements}

This work is supported by the Solar-era.net framework in the project
``PEARL TF-PV'' (Förderkennzeichen: 0324193A) and partly funded by the
HGF project ``Living Lab Energy Campus (LLEC)''.


\printbibliography

\end{document}